\documentclass[fleqn,twoside]{article}
\usepackage{amsmath}
\usepackage{espcrc2}
\usepackage{epsfig}

\usepackage{graphicx}

\newcommand{\pom}{I\!\!P}
\newcommand{\xpom}{x_{\pom}}

\newcommand{\ib}[3]{ibid.\@ #1 (#2) #3}
\newcommand{\PL}[3]{Phys.\@ Lett.\@ #1 (#2) #3}
\newcommand{\NP}[3]{Nucl.\@ Phys.\@ #1 (#2) #3}
\newcommand{\ZP}[3]{Z.\@ Phys.\@ #1 (#2) #3}
\newcommand{\EPJ}[3]{Eur.\@ Phys.\@ J.\@ #1 (#2) #3}
\newcommand{\PR}[3]{Phys.\@ Rev.\@ #1 (#2) #3}

\title{Deep inelastic diffractive scattering at HERA}

\author{Pierre Van Mechelen\address{Universiteit Antwerpen,
        Pierre.VanMechelen@ua.ac.be}%
        \thanks{Postdoctoral fellow of the Fund for Scientific Research - Flanders (Belgium)},}
       
\begin{document}

\begin{abstract}
Recent high precision measurements of deep inelastic diffractive scattering at
HERA are presented in an increased region of phase space.  Current models for
diffractive photon dissociation are compared to the data.
\vspace{1pc}
\end{abstract}

\maketitle

\section{Introduction}

It is well known that soft hadron-hadron scattering can be described by Regge
phenomenology, which models the total, elastic and single and double
diffractive dissociation cross sections in terms of reggeon and pomeron
exchange~\cite{Coll77}.  At high energies, it is the pomeron exchange mechanism
that dominates the total cross section~\cite{Donn92}.  However, until recently,
our understanding of the pomeron in terms of Quantum Chromodynamics (QCD)
\cite{Fors97} remained fragmentary, at best.

The observation of events with a large rapidity gap in the hadronic
final state at HERA~\cite{ZEUS93}, which are attributed to diffractive
dissociation of (virtual) photons, led to a renewed interest in the
study of the underlying dynamics of diffraction.  In deep inelastic
$ep$ scattering, the long hadronic lifetime of the photon at small
Bjorken-$x$ allows to make the link with diffractive dissociation in
hadron-hadron scattering, while the presence of one or more hard scales,
such as the (variable) virtuality $Q^2$, enables perturbative
calculations in QCD.

In the framework of QCD, it is tempting to attribute a partonic
structure to the pomeron~\cite{Inge85}; an approach which proves to be
very successful in describing various aspects of diffractive virtual
photon disscociation.  QCD calculations indicate, however, that the
pomeron does not exhibit a universal behaviour over the full kinematic
range.  Indeed, experimentally a transition is observed when going from
the soft to the hard scattering regime; the effective intercept of the
pomeron trajectory changes from 1.08 in the soft limit to 1.2 in harder
interactions.

Early experimental measurements at HERA of the inclusive cross
section~\cite{H195} led to the conclusion that a partonic pomeron would be
dominated by hard gluons.  This has been confirmed by investigations of the
hadronic final state of diffractive photon dissociation ranging from the study
of inclusive particle spectra to exclusive hard jet and open charm
production~\cite{ZEUS94}.

This article presents recent high precision measurements of the inclusive
diffractive deep inelastic cross section in an increased region of phase
space.  Section~\ref{sec:expdata} presents the experimental data together with
a new NLO QCD fit of the diffractive parton densities.  Current models for
diffractive photon dissociation are subsequently compared to the data in
Sec.~\ref{sec:models}. A summary is given in Sec.~\ref{sec:summary}.

\section{Experimental data}
\label{sec:expdata}

\subsection{Cross section measurements}

In addition to the usual DIS kinematic variables, the photon virtuality
$Q^2$ and the Bjorken scaling variables $x$ and $y$, in the case of
diffractive DIS one introduces the variables $\xpom$ and $\beta$,
defined respectively as the longitudinal momentum fraction of the
proton carried by the colourless exchange causing the rapidity gap, and
the longitudinal momentum fraction of the colourless exchange carried
by the struck quark.  The variable $\beta$ in the $\gamma
\pom$\footnote{$\pom$ is a generic label for the colourless exchange,
which in some models is identified as the pomeron.} collision is
analogous to Bjorken-$x$ in $\gamma p$ interactions.

\begin{figure}
\includegraphics[width=\columnwidth]{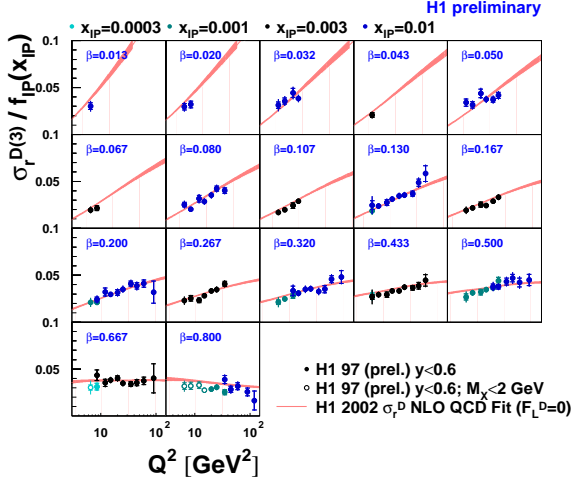}
\caption{The diffractive reduced cross section divided by the pomeron flux as a function
of $Q^2$ in bins of $\beta$ and for different $\xpom$ values.  The band
represents the result of the NLO QCD fit discussed in Sec.~\ref{sec:qcdfit}.}
\label{fig:h1sigmared}
\end{figure}

Figure~\ref{fig:h1sigmared} shows recent measurements obtained by the H1
Collaboration.  The preliminary results are presented as a reduced cross section,
$\sigma_r^{D(3)}$, defined through
\begin{multline}
\frac{d^3\sigma^D}{d\xpom dx dQ^2} = \\
\frac{4\pi\alpha^2}{xQ^4}\left(1-y+\frac{y^2}{2}\right) \times
\sigma_r^{D(3)}(\xpom,x,Q^2)
\end{multline}
and divided by the ``pomeron flux'' $f_{\pom}(\xpom)$. This reduced cross
section is equal to the conventional diffractive structure function,
$F_2^{D(3)}$, up to corrections due the
longitudinal structure function, $F_L^{D(3)}$. The pomeron flux factor is obtained from a
parametrization of the $\xpom$ dependence of the cross section inspired by Regge theory. 
The parametrization works well, as can be seen from the overlap of data points at the same
$\beta$ and $Q^2$ values obtained for different proton momentum losses $\xpom$.  The
intercept of the pomeron trajectory extracted from this data is
\begin{multline}
\alpha_{\pom} = 1.173 \pm 0.018 {\rm\ (stat.)} \\ \pm
0.017 {\rm\ (syst.)} ^{+0.063}_{-0.035} {\rm\ (model)}.
\end{multline}
Positive scaling violations are observed in most of the phase space,
suggesting a large gluon content of the diffractive exchange.  The
ratio of the diffractive to inclusive DIS cross sections is found to be
reasonably flat at fixed $x$ as a function of $Q^2$, indicating that the
same scaling violations occur in both processes.

\begin{figure}
\includegraphics[width=\columnwidth]{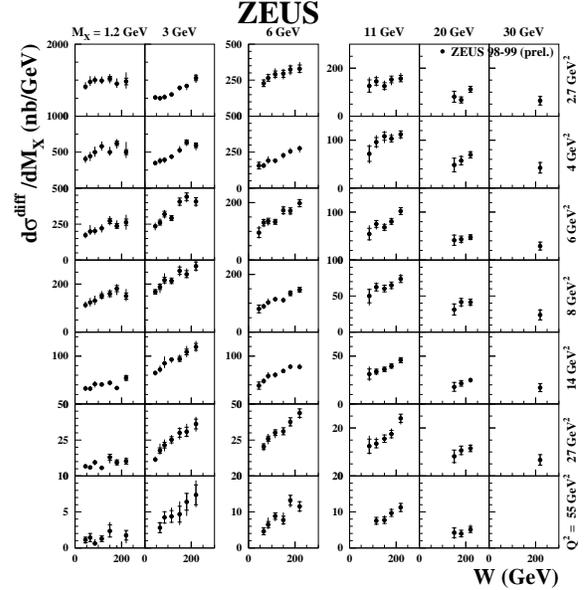}
\caption{The diffractive differential cross section, $d\sigma/dM_X$, as a
function of $W$ in bins of $M_X$ and $Q^2$.}
\label{fig:zeusfpc}
\end{figure}

The ZEUS Collaboration has recently installed a new ``Forward Plug''
calorimeter which covers the range in pseudorapidity $4<\eta<5$ and
increases the measurable range in mass of the proton dissociation
system, $M_Y$, to 2.3 GeV and hence reduces the background due to proton
dissociation.  Preliminary results are shown in Fig.~\ref{fig:zeusfpc}.
It can be observed that in the resonance production region (where the
mass of the photon dissociation system, $M_X$, is smaller than 2 GeV)
there is little dependence on the hadronic mass $W$, while a strong
rise with $W$ is observed at higher $M_X$ for all values of $Q^2$. This
rise is compatible with the energy dependence of inclusive data.

\subsection{Next-to-leading order DGLAP fits}
\label{sec:qcdfit}

Using a parametrization based on Chebychev polynomials at a starting
scale of $Q^2_0 = 3 {\rm\ GeV}^2$, quark and gluon densities have been
fitted to the observed H1 cross section by applying the DGLAP evolution
equations.  Subleading reggeon exchanges are included assuming the
structure function of the pion. The fit, which includes the data shown
in Fig.~\ref{fig:h1sigmared} together with data at higher $Q^2$ shown
in Fig.~\ref{fig:sci}, yields $\chi^2/ndf = 308.7/306$.

\begin{figure}
\includegraphics[width=\columnwidth]{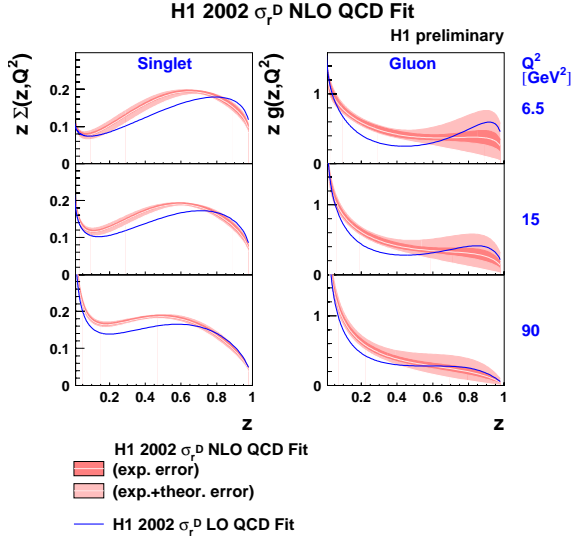}
\caption{Diffractive quark singlet ($6 * u$ with
$u=d=s=\overline{u}=\overline{d}=\overline{s}$) and gluon density.  The pomeron
flux is normalized to unity at $\xpom=0.003$.}
\label{fig:h1fits}
\end{figure}

Figure~\ref{fig:h1fits} shows the result of the next-to-leading order (NLO) QCD fit, with full propagation of
statistical, systematic experimental and theoretical errors. The momentum fraction carried by gluons is estimated to be
$75 \pm 15\% 
$.

\section{Model comparisons}
\label{sec:models}

\subsection{The Soft Colour Interaction model at large \boldmath{$Q^2$}}

A model that does not use the notion of a pomeron at all is the Soft
Colour Interaction (SCI) model~\cite{Edin96}.  Instead, the standard
matrix element and parton shower description of deep inelastic
scattering is used, which, at low Bjorken-$x$, is dominated by
boson-gluon fusion. In addition, non-perturbative interactions affect
the final colour connections between partons, while leaving the parton
momenta unchanged.  This may result in an interruption of the colour
strings between partons, thus creating a large rapidity gap in the
final state.  The probability for such a SCI has to be fixed by the
experimental data. 

\begin{figure}
\includegraphics[width=\columnwidth]{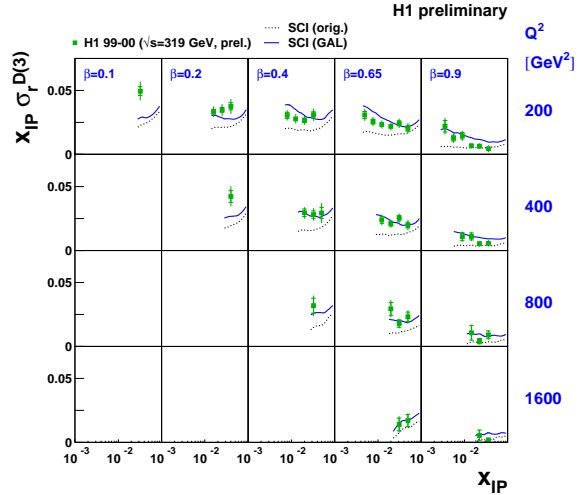}
\caption{The diffractive reduced cross section, multiplied by $\xpom$, is shown
as a function of $\xpom$ in bins of $\beta$ and $Q^2$.  The curves represent the
SCI model with and without the GAL.}
\label{fig:sci}
\end{figure}

Figure~\ref{fig:sci} compares the SCI model to recently obtained data at large
$Q^2$. While the application of a ``Generalized Area Law'' (GAL), which favours
soft colour interactions that make the colour strings between partons shorter,
clearly improves the description of the data (as is also observed at lower
$Q^2$), systematic deviations occur towards low values of $\beta$.

\subsection{The Saturation model at low \boldmath{$Q^2$}}

In this and the subsequent model, the virtual photon is considered to
fluctuate into partonic configurations, e.g.\@ $q\overline{q}$ or
$q\overline{q}g$.  In the proton rest frame and at low Bjorken-$x$,
this happens long before the actual interaction with the proton.  Both
states, $q\overline{q}$ and $q\overline{q}g$, can be described by
dipole wave-functions \cite{Bart99}.

\begin{figure}
\includegraphics[width=\columnwidth]{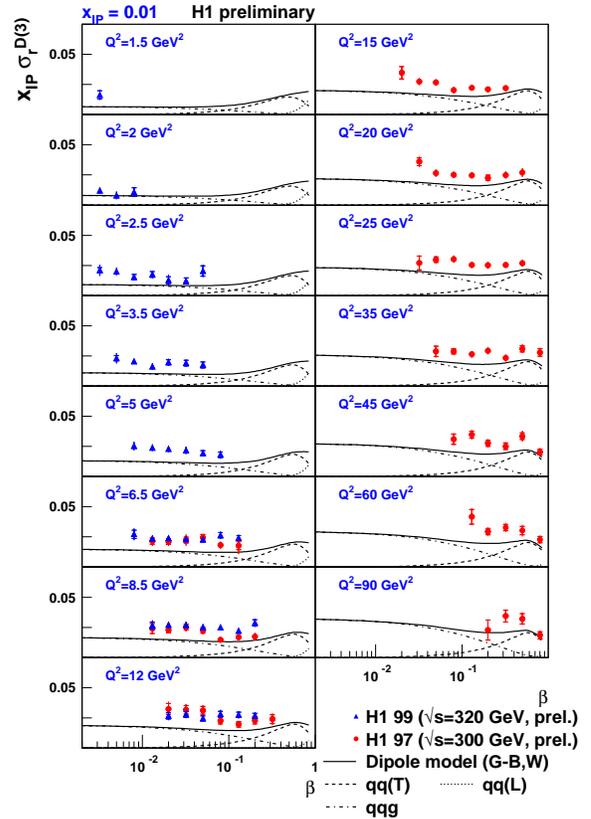}
\caption{The diffractive reduced cross section, multiplied by $\xpom$, is shown
as a function of $\beta$ at $\xpom=0.01$ and in bins of $Q^2$. The curves
represent the prediction of the Saturation model decomposed in longitudinal and transverse
$q\overline{q}$ and $q\overline{q}g$ contributions.}
\label{fig:sat}
\end{figure}

The Saturation model~\cite{Gole99} proposes a para\-metrization for the dipole-proton cross
section:
\begin{equation}
\sigma(x,r^2) = \sigma_0 \left[1 -
\exp\left(-\frac{r^2}{4R_0^2(x)}\right)\right],
\end{equation}
with
\begin{equation}
R_0(x) = \frac{1}{{\rm GeV}}\left(\frac{x}{x_0}\right)^{\lambda/2};
\end{equation}
$x_0$ and $\lambda$ are parameters of the model.  This parameterization exhibits colour transparency for small dipole sizes, $r$, and
saturation of the cross section towards low $x$ and low $Q^2$.

Figure~\ref{fig:sat} shows that this model consistently lies beneath the
measured data points.

\subsection{Two-gluon exchange model}

This QCD model~\cite{Bart96} describes the elastic scattering of $q\overline{q}$ and 
$q\overline{q}g$ states off the proton through the exchange of two gluons in a
net colour-singlet configuration.  The perturbative calculation requires large
transverse momenta of all outgoing partons and $\xpom < 0.01$ to avoid the
valence quark region in the proton.

\begin{figure*}
\begin{center}
\includegraphics*[width=0.7\textwidth]{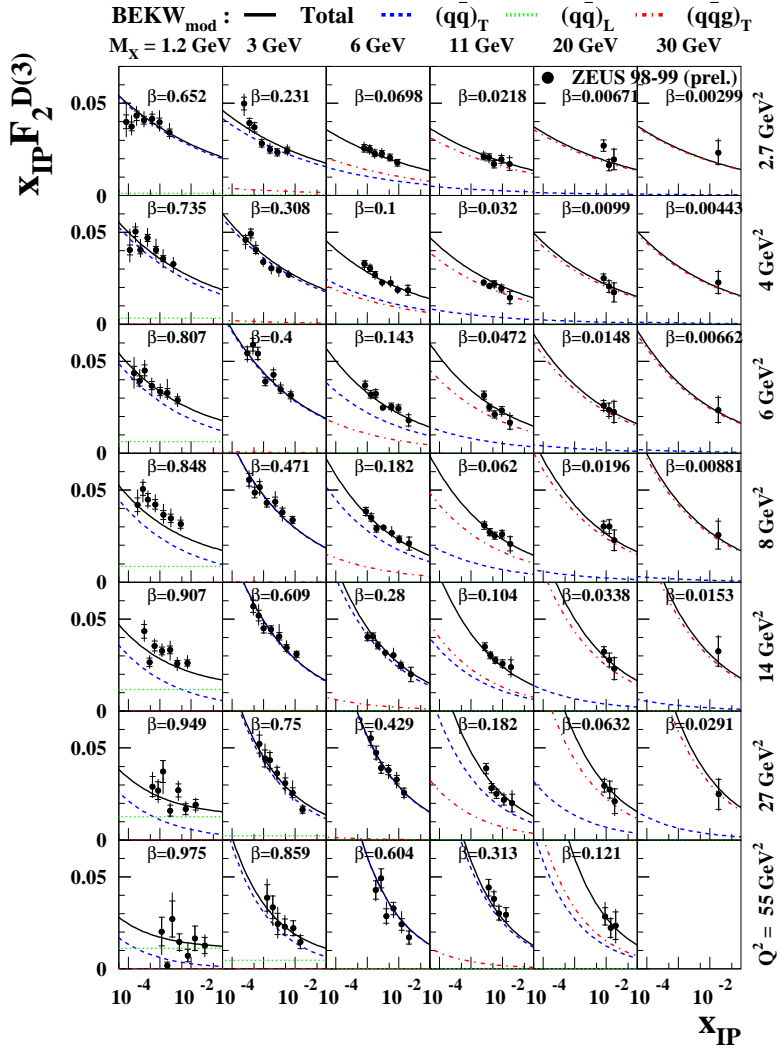}
\end{center}
\caption{The diffractive structure function, multiplied by $\xpom$, is shown as
a function of $\xpom$ in bins of $\beta$ and $Q^2$.  The curves represent the
prediction of the
2-gluon exchange model decomposed in longitudinal and transverse
$q\overline{q}$ and $q\overline{q}g$ contributions.}
\label{fig:2gluon}
\end{figure*}

Despite these restrictions, Fig.~\ref{fig:2gluon} shows that the model describes
the data rather well.

\section{Summary}
\label{sec:summary}

High precision measurements of the diffractive reduced cross section in deep
inelastic scattering have been performed by the H1 and ZEUS collaborations in
an increased region of phase space. The data support Regge factorization
(provided subleading trajectories can contribute) with a value for the pomeron
intercept which is larger than for the soft pomeron.  New NLO QCD fits of the
diffractive parton densities are available and can be used to test QCD hard
scattering factorization.  Models describe the
data in general, but some new discrepancies are uncovered with the increased
accessible phase space.

\section*{Acknowledgments}
I am indebted to all members of the H1 and ZEUS Collaborations who contributed
to these results by collecting and analyzing the experimental data.

\end{document}